\documentclass{edm_template}
\begin{document}


\title{DAS3H: Modeling Student Learning and Forgetting for Optimally Scheduling Distributed Practice of Skills}

%
%
%
%
%

\numberofauthors{4} 
%
\author{
  \begin{tabular}[t]{p{2.2\auwidth}}\centering
Benoît~Choffin, Fabrice Popineau, Yolaine Bourda\\
    \affaddr{LRI/CentraleSupélec -- University of Paris-Saclay}\\
    \affaddr{Gif-sur-Yvette, France}\\
    \email{\{benoit.choffin, fabrice.popineau, yolaine.bourda\}@lri.fr}
\and
\alignauthor
Jill-Jênn Vie\\
    \affaddr{RIKEN AIP}\\
    \affaddr{Tokyo, Japan}\\
    \email{vie@jill-jenn.net}\\
\end{tabular}
}

\maketitle
\begin{abstract}


Spaced repetition is among the most studied learning strategies in the cognitive science literature. It consists in temporally distributing exposure to an information so as to improve long-term memorization. Providing students with an adaptive and personalized distributed practice schedule would benefit more than just a generic scheduler. However, the applicability of such adaptive schedulers seems to be limited to pure memorization, e.g. flashcards or foreign language learning. In this article, we first frame the research problem of optimizing an adaptive and personalized spaced repetition scheduler when memorization concerns the application of underlying multiple skills. To this end, we choose to rely on a student model for inferring knowledge state and memory dynamics on any skill or combination of skills. We argue that no knowledge tracing model takes both memory decay and multiple skill tagging into account for predicting student performance. As a consequence, we propose a new student learning and forgetting model suited to our research problem: DAS3H builds on the additive factor models and includes a representation of the temporal distribution of past practice on the skills involved by an item. In particular, DAS3H allows the learning and forgetting curves to differ from one skill to another. Finally, we provide empirical evidence on three real-world educational datasets that DAS3H outperforms other state-of-the-art EDM models. These results suggest that incorporating both item-skill relationships and forgetting effect improves over student models that consider one or the other.

\end{abstract}

%

\keywords{Student modeling, adaptive spacing, memory, knowledge components, q-matrix, optimal scheduling}

\section{Introduction}




Learners have to manage their studying time wisely: they constantly have to make a trade-off between acquiring new knowledge and reviewing previously encountered learning material. Considering that learning often involves building on old knowledge (e.g. in mathematics) and that efforts undertaken in studying new concepts may be significant, this issue should not be taken lightly. However, only few school incentive structures encourage long-term retention, making students often favor short-term memorization and poor learning practices \cite{roediger2018remembering,mozer2016predicting}.

Fortunately, there are simple learning strategies that help students efficiently manage their learning time and improve long-term memory retention at a small cost. Among them, the \textit{spacing} and the \textit{testing} effects have been widely replicated \cite{roediger2011intricacies, cepeda2008spacing} since their discovery in the 19\textsuperscript{th} century. Both of them are recommended by cognitive scientists \cite{pashler2007organizing, weinstein2018teaching} in order to improve public instruction. The spacing effect states that temporally distributing learning episodes is more beneficial to long-term memory than learning in a single massed study session. The testing effect \cite{roediger2006test, carpenter2008effects} -- also known as \textit{retrieval practice} -- basically consists in self-testing after being exposed to new knowledge instead of simply reading the lesson again. This test can take multiple forms: free recall, cued recall, multiple-choice questions, application exercises, and so on. A recent meta-analysis on the testing effect \cite{adesope2017rethinking} found a strong and positive overall effect size of $g = 0.61$ for testing compared to non-testing reviewing strategies. Another meta-analysis \cite{pan2018transfer} investigated whether learning with retrieval practice could transfer to different contexts and found a medium yet positive overall transfer effect size of $d=0.40$. Combining both strategies is called \textit{spaced retrieval practice}: temporally distributing tests after a first exposure to knowledge.

Recent research effort has been put on developing adaptive and personalized spacing schedulers for improving long-term retention of flashcards \cite{tabibian2019enhancing, reddy2017accelerating, lindsey2014improving}. Compared to non-adaptive schedulers, they show substantial improvement of the learners' retention at immediate and delayed tests \cite{mettler2016comparison}. However, and to the best of our knowledge, there is no work on extending these algorithms when knowledge to be remembered concerns the application of underlying skills. Yet, the spacing effect is not limited to vocabulary learning or even pure memorization: it has been successfully applied to the acquisition and generalization of abstract science concepts \cite{vlach2012distributing} and to the practice of mathematical skills in a real educational setting \cite{barzagar2019distributing}. Conversely, most models encountered in knowledge tracing involve multiple skills, but do not model forgetting. The goal of the present article is to start filling this gap by developing a student learning and forgetting model for inferring skills knowledge state and memory dynamics. This model will serve as a basis for the future development of adaptive and personalized skill practice scheduling algorithms for improving learners' long-term memory.

Our contribution is two-fold. We first frame our research problem for extending the flashcards-based adaptive spacing framework to contexts where memorization concerns the application of underlying skills. In that perspective, students learn and reinforce skill mastery by practicing items involving that skill. We argue that this extension requires new student models to model learning and forgetting processes when multiple skills are involved by a single item. Thus, we also propose a new student model, coined DAS3H, that extends DASH \cite{lindsey2014improving,mozer2016predicting} and accounts for memory decay and the benefits of practice when an item can involve multiple knowledge components. Finally, we provide empirical evidence on three publicly available datasets showing that our model outperforms other state-of-the-art student models.

\section{Related work}
\label{related_works}

In this section, we first detail related work on adaptive spacing algorithms before turning to student modeling.

In what follows, we will index students by $s \in \llbracket 1,S \rrbracket$, items (or questions, exercises) by $j \in \llbracket 1,J \rrbracket$, skills or knowledge components (KCs) by $k \in \llbracket 1,K \rrbracket$, and timestamps by $t \in \mathbb{R}^+$ (in days). To be more convenient, we assume that timestamps are encoded as the number of days elapsed since the first interaction with the system. It is sufficient because we only need to know the duration between two interactions. $Y_{s,j,t} \in \{0,1\}$ gives the binary correctness of student $s$ answering item $j$ at time $t$. $\sigma$ is the logistic function: $\forall x \in \mathbb{R}, \sigma(x)=1/(1+\exp(-x))$. $KC(.)$ takes as input an item index $j$ and outputs the set of skill indices involved by item $j$.

Let us quickly detail what we mean by \textit{skill}. In this article, we assimilate skills and knowledge components. Knowledge components are atomistic components of knowledge by which items are tagged. An item may have one or more KCs, and this information is synthesized by a so-called binary q-matrix \cite{tatsuoka1983rule}: $\forall (j,k) \in \llbracket 1,J \rrbracket \times \llbracket 1,K \rrbracket, \: q_{jk}=\mathds{1}_{k \in KC(j)}$. We assume that the probability of answering correctly an item $j$ that involves skill $k$ depends on the student's mastery of skill $k$; conversely, we measure skill mastery by the ability of student $s$ to remember skill $k$ and apply it to solve any (possibly unseen) item that involves skill $k$.

\subsection{Adaptive spacing algorithms}
Adaptive spacing schedulers leverage the spaced retrieval learning strategy to maximize learning and retention of a set of items. They proceed by sequentially deciding which item to ask the user at any time based on the user's past study history. Items to memorize are often represented by flashcards, i.e. cards on which one side contains the question (e.g. \textit{When did the Great Fire of London occur?} or \textit{What is the correct translation of ``manger'' in English?}) and the other side contains the answer.

Early adaptive spacing systems made use of physical flashcards \cite{leitner1972so} but the advent of computer-assisted instruction made possible the development of electronic flashcards \cite{wozniak1994optimization}, thus allowing more complex and personalized strategies for optimal reviewing. Nowadays, several adaptive spacing softwares are available to the general public, e.g. Anki\footnote{\url{https://apps.ankiweb.net/}}, SuperMemo\footnote{\url{https://www.supermemo.com/}}, and Mnemosyne\footnote{\url{https://mnemosyne-proj.org/}}.

Originally, adaptive reviewing systems took decisions based on heuristics and handmade rules \cite{leitner1972so, pimsleur1967memory, wozniak1994optimization}. Though maybe effective in practice \cite{metzler2009does}, these early systems lack performance guarantees \cite{tabibian2019enhancing}. Recent research works started to tackle this issue: for instance, Reddy et al. propose a mathematical formalization of the Leitner system and a heuristic approximation used for optimizing the review schedule \citeyear{reddy2016unbounded}.

A common approach for designing spaced repetition adaptive schedulers consists in modeling human memory statistically and recommending the item whose memory strength is closest to a fixed value $\theta$ \cite{mozer2016predicting,lindsey2014improving,metzler2009does}. Khajah, Lindsey and Mozer found that this simple heuristic is only slightly less efficient than exhaustive policy search in many situations \citeyear{khajah2014maximizing}. It has the additional advantage to fit into the notion of ``desirable difficulties'' coined by Bjork \cite{bjork1994memory}. Pavlik and Anderson \cite{pavlik2008using} use an extended version of ACT-R declarative memory model to build an adaptive scheduler for optimizing item practice (in their case, Japanese-English word pairs) given a limited amount of time. ACT-R is originally capable of predicting item correctness and speed of recall by taking recency and frequency of practice into account. Pavlik and Anderson extend ACT-R to capture the spacing effect as well as item, learner, and item-learner interaction variability. The adaptive scheduler uses the model estimation of memory strength gain at retention test per unit of time to decide when to present each pair of words to a learner.

Other approaches do not rely on any memory model: Reddy, Levine and Dragan formalize this problem as a POMDP (Partially Observable Markov Decision Process) and approximately solve it within a deep reinforcement learning architecture \citeyear{reddy2017accelerating}. However, they only test their algorithm on simulated students. A more recent work \cite{tabibian2019enhancing} formalizes the spaced repetition problem with marked temporal point processes and solves a stochastic optimal control problem to optimally schedule spaced review of items. Mettler, Massey and Kellman \cite{mettler2016comparison} compare an adaptive spacing scheduler (ARTS) to two fixed spacing conditions. ARTS leverages students' response times, performance, and number of trials to dynamically compute a priority score for adaptively scheduling item practice. Response time is used as an indicator of retrieval difficulty and thus, learning strength.

Our work can more generally relate to the problem of automatic optimization of teaching sequences. Rafferty et al. formulate this problem as a POMDP planning problem \citeyear{rafferty2011faster}. Whitehill and Movellan build on this work but use a hierarchical control architecture for selecting optimal teaching actions \citeyear{whitehill2018approximately}. Lan et al. use contextual bandits to select the best next learning action by using an estimation of the student's knowledge profile \citeyear{lan2016contextual}. Many intelligent tutoring systems (ITS) use mastery learning within the Knowledge Tracing \cite{corbett1994knowledge} framework: making students work on a given skill until the system infers that they have mastered it.

We can see that the traditional adaptive spacing framework already uses a spaced retrieval practice strategy to optimize the student's learning time. However, it is not directly adapted to learning and memorization of skills. In this latter case, specific items are the only way to practice one or multiple skills, because we do not have to memorize the content directly. Students who master a skill should be able to generalize to unseen items that also involve that skill. In Section \ref{pb_form}, we propose an extension of this original framework in order to apply adaptive spacing algorithms to the memorization of skills.

\subsection{Modeling student learning and forgetting}

The history of scientific literature on student modeling is particularly rich. In what follows, we focus on the subproblem of modeling student learning and forgetting based on student performance data.

As Vie and Kashima recall \citeyear{Vie2019}, two main approaches have been used for modeling student learning and predicting student performance: Knowledge Tracing and Factor Analysis.

Knowledge Tracing \cite{corbett1994knowledge} models the evolution of a student's knowledge state over time so as to predict a sequence of answers. The original and still most widespread model of Knowledge Tracing is Bayesian Knowledge Tracing (BKT). It is based on a Hidden Markov Model where the knowledge state of the student is the latent variable and skill mastery is assumed binary. Since its creation, it has been extended to overcome its limits and account for instance for individual differences between students \cite{yudelson2013individualized}. More recently, Piech et al. replaced the original Hidden Markov Model framework with a Recurrent Neural Network and proposed a new Knowledge Tracing model called Deep Knowledge Tracing (DKT) \citeyear{piech2015deep}. Despite a mild controversy concerning the relevance of using deep learning in an educational setting \cite{wilson2016estimating}, recent works continue to develop this line of research \cite{zhang2017dynamic,minn2018deep}.

Contrary to Knowledge Tracing, Factor Analysis does not originally take the order of the observations into account. IRT (Item Response Theory) \cite{van2013handbook} is the canonical model for Factor Analysis. In its simplest form, IRT reads:
\[\mathbb{P}(Y_{s,j} = 1) = \sigma(\alpha_s - \delta_j)\]
with $\alpha_s$ ability of student $s$ and $\delta_j$ difficulty of item $j$. One of the main assumptions of IRT is that the student ability is static and cannot change over time or with practice. Despite its apparent simplicity, IRT has proven to be a robust and reliable EDM model, even outperforming much more complex architectures such as DKT \cite{wilson2016back}. IRT can be extended to represent user and item biases with vectors instead of scalars. This model is called MIRT, for Multidimensional Item Response Theory:
\[\mathbb{P}(Y_{s,j} = 1) = \sigma \left(\langle \alpha_s , \delta_j \rangle + d_j\right).\]
In this case, $\alpha_s$ and $\delta_j$ are $d$-dimensional vectors, and $d_j$ is a scalar that captures the easiness of item $j$. $\langle., .\rangle$ is the usual dot product between two vectors.

More recent works incorporated temporality in Factor Analysis models, by taking practice history into account. For instance, AFM (Additive Factor Model) \cite{cen2006learning} models:
\[\mathbb{P}(Y_{s,j} = 1) = \sigma \left(\sum_{k \in KC(j)} \beta_k + \gamma_k a_{s,k}\right)\]
with $\beta_k$ easiness of skill $k$ and $a_{s,k}$ number of attempts of student $s$ on KC $k$ prior to this attempt. Performance Factor Analysis \cite{pavlik2009performance} (PFA) builds on AFM and uses past outcomes of practice instead of simple encounter counts:
\[\mathbb{P}(Y_{s,j} = 1) = \sigma \left(\sum_{k \in KC(j)} \beta_k + \gamma_k c_{s,k} + \rho_k f_{s,k}\right)\]
with $c_{s,k}$ number of correct answers of student $s$ on KC $k$ prior to this attempt and $f_{s,k}$ number of wrong answers of student $s$ on KC $k$ prior to this attempt.

Ekanadham and Karklin take a step further to account for temporality in the IRT model and extend the two-parameter ogive IRT model (2PO model) by modeling the evolution of the student ability as a Wiener process \citeyear{ekanadham2017t}. However, they do not explicitly account for student memory decay.

The recent framework of KTM (Knowledge Tracing Machines) \cite{Vie2019} encompasses several EDM models, including IRT, MIRT, AFM, and PFA. KTMs are based on factorization machines and model the probability of correctness as follows:
\[\mathbb{P}(Y_t = 1) = \sigma\left(\mu + \sum_{i=1}^{N} w_i x_{t,i} + \sum_{1 \leq i < \ell \leq N} x_{t,i} x_{t,\ell} \langle v_i, v_\ell \rangle \right)\]
where $\mu$ is a global bias, $N$ is the number of abstract features, be it item parameters, temporal features, etc., $x_t$ is a sample gathering all features collected at time $t$: which student answers which item, and information regarding prior attempts, $w_i$ is the bias of feature $i$ and $v_i \in \mathbb{R}^d$ its embedding. The features involved in a sample $x_t$ are typically in sparse number, so this probability can be computed efficiently. In KTM, one can recover several existing EDM models by selecting the appropriate features to consider in the modeling. For instance, if we consider user and item features only, we recover IRT. If we consider the skill features in the q-matrix, and the counter of prior successes and failures at skill level, we recover PFA.

One of the very first works on human memory modeling dates back to 1885 and stems from Ebbinghaus \cite{ebbinghaus2013memory}. He models the probability of recall of an item as an exponential function of memory strength and delay since last review. More recently, Settles and Meeder propose an extension of the original exponential forgetting curve model, the half-life regression  \citeyear{settles2016trainable}. They estimate item memory strength as an exponential function of a set of features that contain information on the past practice history and on the item to remember (lexeme tag features, in their case). More sophisticated memory models have also been proposed: for instance ACT-R (Adaptive Character of Thought–Rational) \cite{anderson1997act} and MCM (Multiscale Context Model) \cite{pashler2009predicting}.

Walsh et al. \cite{walsh2018mechanisms} offer a comparison of three computational memory models: ACT-R declarative memory model \cite{pavlik2008using}, Predictive Performance Equation (PPE) and a generalization of Search of Associative Memory (SAM). These models differ in how they predict the impact of spacing on subsequent relearning, after a long retention interval. PPE is the only one to predict that spacing may accelerate subsequent relearning (``\textit{spacing accelerated relearning}'') -- an effect that was empirically underlined by their experiment. PPE showed also superior fit to experimental data, compared to SAM and ACT-R.

DASH \cite{mozer2016predicting,lindsey2014improving} bridges the gap between factor analysis and memory models. DASH stands for \underline{D}ifficulty, \underline{A}bility, and \underline{S}tudent \underline{H}istory. Its formulation reads:
\[\mathbb{P}\left(Y_{s,j,t}=1\right)=\sigma(\alpha_s - \delta_j + h_{\theta}(\mathrm{t}_{s,j,1:l},\mathrm{y}_{s,j,1:l-1}))\]
with $h_\theta$ a function parameterized by $\theta$ (learned by DASH) that summarizes the effect of the $l-1$ previous attempts where student $s$ reviewed item $j$ ($\mathrm{t}_{s,j,1:l-1}$) and the binary outcomes of these attempts ($\mathrm{y}_{s,j,1:l-1}$). Their main choice for $h_{\theta}$ is:
\begin{align*}
    h_{\theta}(\mathrm{t}_{s,j,1:l},\mathrm{y}_{s,j,1:l-1}) = \sum_{w=0}^{W-1} & \theta_{2w+1}\log(1+c_{s,j,w}) \\
    &- \theta_{2w+2}\log(1+a_{s,j,w})
\end{align*}
with $w$ indexing a set of expanding time windows, $c_{s,j,w}$ is the number of correct outcomes of student $s$ on item $j$ in time window $w$ out of a total of $a_{s,j,w}$ attempts. The time windows $w$ are not disjoint and span increasing time intervals. They allow DASH to account for both learning and forgetting processes. The use of log counts induces diminishing returns of practice inside a given time window and difference of log counts formalizes a power law of practice. The time module $h_{\theta}$ is inspired by ACT-R \cite{anderson1997act} and MCM \cite{pashler2009predicting} memory models.

We can see that Lindsey et al. \cite{lindsey2014improving} make use of the additive factor models framework for taking memory decay and the benefits of past practice into account. Their model outperforms IRT and a baseline on their dataset COLT, with an accumulative prediction error metric. To avoid overfitting and making model training easier, they use a hierarchical Bayesian regularization.

To the best of our knowledge, no knowledge tracing model accounts for both multiple skills tagging \textit{and} memory decay. We intend to bridge this gap by extending DASH.

\section{Framing the problem}
\label{pb_form}

In our setting, the student learns to master a set of skills by sequentially interacting with an adaptive spacing system. At each iteration, this system selects an item (or exercise, or question) for the student, e.g. \textit{What is $\lim_{x \rightarrow 0} (\sin{x})/x$?}. This selection is made by optimizing a utility function $l$ that rewards long-term mastery of the set of KCs to learn. Then, the student answers the item and the system uses the correctness of the answer to update its belief concerning the student memory and learning state on the skills involved by the item. Finally, the system provides the student a corrective feedback.

In a nutshell, our present research goal is to maximize mastery and memory of a fixed set of skills among students during a given time interval while minimizing the time spent studying.

We rely on the following assumptions:
\begin{itemize}
    \item information to learn and remember consists in a set of skills\footnote{These skills may be organized into a graph of prerequisites, but this goes beyond the scope of this article.} $k \in \llbracket1,K\rrbracket$;
    \item skill mastery and memorization of student $s$ at time $t$ is measured by the ability of $s$ to answer an (unseen) item involving that skill, i.e. by their ability to generalize to unseen material;
    \item students first have access to some theoretical knowledge about skills, but learning happens with retrieval practice;
    \item items are tagged with one or multiple skills and this information is synthesized inside a binary q-matrix \cite{tatsuoka1983rule};
    \item students forget: skill mastery decreases as time goes by since last practice of that skill;
\end{itemize}

Unlike Lindsey et al. \citeyear{lindsey2014improving}, we do not assume that items involving skill $k$ are interchangeable: their difficulties, for instance, may differ from one another. Thus, the selection phase is two-fold in that it requires to select the skill to practice \textit{and} the item to present. In theory, there should be at least one item for practicing every skill $k$; in practice, one item would be too few, since the student would probably ``overfit'' on the item. This formalization easily encompasses the flashcards-based adaptive spacing framework: it only requires to associate every item with a distinct skill. This wipes out the need to select an item after the skill.

Different utility functions $l$ can be considered. For instance, Reddy, Levine and Dragan consider both the likelihood of recalling all items and the expected number of items recalled \citeyear{reddy2017accelerating}. In our case, the utility function should account for the uncertainty of future items to answer. Indeed, if the goal of the user is to prepare for an exam, the system must take into account that the user will probably have to answer items that they did not train with.

To tackle this problem, like previous work \cite{mozer2016predicting,lindsey2014improving}, we choose to rely on a student learning and forgetting model. In our case, this model must be able to quantify mastery and memory for any skill or combination of skills. In the next section, we present our main contribution: a new student learning and forgetting model, coined DAS3H.

\section{Our model DAS3H}

\begin{sloppypar} We now describe our new student learning and forgetting model: DAS3H stands for item \underline{D}ifficulty, student \underline{A}bility, \underline{S}kill, and \underline{S}tudent \underline{S}kill practice \underline{H}istory, and builds on the DASH model presented in Section \ref{related_works}. Lindsey et al.\;\cite{lindsey2014improving} show that DASH outperforms a hierarchical Bayesian version of IRT on their experimental data, which consist in student-item interactions on a flashcard-based foreign (Spanish) language vocabulary reviewing system. They already talk about knowledge components, but they use this concept to cluster similar words together (e.g. all conjugations of a verb). Thus, in their setting, an item has exactly one knowledge component; different items can belong to the same knowledge component if they are close enough. As a consequence, their model formulation does not handle multiple skills item tagging, which is common in other disciplines such as in mathematics. Moreover, they assume that the impact of past practice on the probability of correctness does not vary across the skills: indeed, DASH has only two biases per time window $w$, $\theta_{2w+1}$ for past wins and $\theta_{2w+2}$ for past attempts. It may be a relevant assumption to prevent overfitting when the number of skills is high, but at the same time it may degrade performance when the set of skills is very diverse and inhomogeneous.\end{sloppypar}

DAS3H extends DASH to items with multiple skills, and allows the influence of past practice on present performance to differ from one skill to another. One could argue that we could aggregate every existing combination of skills into a distinct skill to avoid the burden of handling multiple skills. However, this solution would not be satisfying since the resulting model would for instance not be able to capture item similarities between two items that share all but one skill in common. The use of a representation of multiple skills allows to account for knowledge transfer from one item to another. The item-skill relationships are usually synthesized by a q-matrix and generally require domain experts' labor.

We also leverage the recent Knowledge Tracing Machines framework \cite{Vie2019} to enrich the DASH model by embedding the features in $d$ dimensions and model pairwise interactions between those features. So far, KTMs have not been tried with memory features.

In brief, we extend DASH in three ways:
\vspace{-4mm}
\begin{itemize}
    \item Extension to handle multiple skills tagging: new temporal module $h_{\theta}$ that also takes the multiple skills into account. The influence of the temporal distribution of past practice and of the outcomes of these previous attempts may differ from one skill to another;
    \item Estimation of easiness parameters for \textit{each} item $j$ and skill $k$;
    \item Use of KTMs \cite{Vie2019} instead of mere logistic regression.
\end{itemize}

For an embedding dimension of $d=0$, the quadratic term of KTM is cancelled out and our model DAS3H reads:
\begin{align*}
    \mathbb{P}\left(Y_{s,j,t}=1\right)=\sigma (&\alpha_s - \delta_j + \sum_{k \in KC(j)} \beta_k +\\ &+h_{\theta}\left(\mathrm{t}_{s,j,1:l},\mathrm{y}_{s,j,1:l-1}\right)).
\end{align*}
Following Lindsey et al. \citeyear{lindsey2014improving}, we choose:
\begin{align*}
    h_{\theta}(\mathrm{t}_{s,j,1:l},\mathrm{y}_{s,j,1:l-1}) = \sum_{k \in KC(j)}&\sum_{w=0}^{W-1}\theta_{k,2w+1}\log(1+c_{s,k,w})\\
    &- \theta_{k,2w+2}\log(1+a_{s,k,w}).
\end{align*}
Thus, the probability of correctness of student $s$ on item $j$ at time $t$ depends on their ability $\alpha_s$, the difficulty of the item $\delta_j$ and the sum of the easiness $\beta_k$ of the skills involved by item $j$. It also depends on the temporal distribution and the outcomes of past practice, synthesized by $h_{\theta}$. In $h_{\theta}$, $w$ denotes the index of the time window, $c_{s,k,w}$ denotes the amount of times that KC $k$ has been correctly recalled in window $w$ by student $s$ earlier, $a_{s,k,w}$ the amount of times that KC $k$ has been encountered in time window $w$ by student $s$ earlier. Intuitively, $h_{\theta}$ can be seen as a sum of memory strengths, one for each skill involved in item $j$.

For higher embedding dimensions $d > 0$, in our implementation we use probit as the link function. All features are embedded in $d$ dimensions and their interaction is modeled in a pairwise manner. For a more thorough description of KTMs, see \cite{Vie2019}. To implement a model within the KTM framework, one must decide which features to encode in the sparse $x$ vector. In our case, we chose user $s$, item $j$, skills $k \in KC(j)$, wins $c_{s,k,w}$ and attempts $a_{s,k,w}$ for each time window $w$.

Compared to DASH and if we forget about additional parameters induced by the regularization scheme, DAS3H has $(d+1)(K+2W(K-1))$ more feature parameters to estimate. To avoid overfitting, we use additional hierarchical distributional assumptions for the parameters to estimate, as described in the next section.

\section{Experiments}

To evaluate the performance of our model, we compared DAS3H to several state-of-the-art student models on three different educational datasets. These models have been detailed in Section \ref{related_works}.

\subsection{Experimental setting}
We perform 5-fold cross-validation at the student level for our experiments. This means that the student population is split into 5 disjoint groups and that cross-validation is made on this basis. This evaluation method, also used in \cite{Vie2019}, has the advantage to show how well an educational data mining model generalizes over previously unseen students.

Following previous work \cite{rendle2012factorization,Vie2019} we use hierarchical distributional assumptions when $d>0$ to help model training and avoid overfitting. More precisely, each feature weight and feature embedding component follows a normal prior distribution $\mathcal{N}(\mu,1/\lambda)$ where $\mu$ and $\lambda$ follow hyperpriors $\mu \sim \mathcal{N}(0,1)$ and $\lambda \sim \Gamma(1,1)$. In their article \cite{lindsey2014improving}, Lindsey et al. took a similar approach but they assumed that the $\alpha_s$ and the $\delta_i$ followed different distributions. Contrary to us, they did not regularize the parameters $\theta_w$ associated with the practice history of a student: our situation is different because we have more parameters to estimate than them. We use the same time windows as Lindsey et al. \cite{lindsey2014improving}: $\{1/24,1,7,30,+\infty\}$. Time units are expressed in days.

Our models were implemented in Python. Code for replicating our results is freely available on Github\footnote{\url{https://github.com/BenoitChoffin/das3h}}. Like Vie and Kashima \citeyear{Vie2019}, we used \texttt{pywFM}\footnote{\url{https://github.com/jfloff/pywFM}} as wrapper for \texttt{libfm}\footnote{\url{http://libfm.org/}} \cite{rendle2012factorization} for models with $d>0$. We used 300 iterations for the MCMC Gibbs sampler. When $d=0$, we used the \texttt{scikit-learn} \cite{pedregosa2011scikit} implementation of logistic regression with L2 regularization.

\begin{table*}
\centering
\begin{tabular}{@{}lrrrrrrrr@{}}
\toprule
Dataset & Users & Items & Skills & Interactions & \makecell[cl]{Mean \\ correctness} & 
\makecell[cl]{Skills \\ per item} &
\makecell[cl]{Mean \\ skill delay} &
\makecell[cl]{Mean \\ study period} \\
\midrule
assist12 & 24,750 & 52,976 & 265 & 2,692,889 & 0.696 & 1.000 & 8.54 & 98.3 \\
bridge06 & 1,135 & 129,263 & 493 & 1,817,427 & 0.832 & 1.013 & 0.83 & 149.5 \\
algebra05 & 569 & 173,113 & 112 & 607,000 & 0.755 & 1.363 & 3.36 & 109.9 \\
\bottomrule
\end{tabular}
\caption{Datasets characteristics}
\label{data_caracs}
\end{table*}
\begin{table*}
\centering
\parbox{.42\linewidth}{%
\centering
\begin{tabular}{cccc}
\toprule
                      model &  dim &             AUC $\uparrow$ & NLL $\downarrow$\\
\midrule
DAS3H & 0 & $\textbf{0.826} \pm 0.003$ & $\textbf{0.414} \pm 0.011$ \\
DAS3H & 5 & $\textbf{0.818} \pm 0.004$ & $\textbf{0.421} \pm 0.011$ \\
DAS3H & 20 & $\textbf{0.817} \pm 0.005$ & $\textbf{0.422} \pm 0.007$ \\
DASH & 5 & $0.775 \pm 0.005$ & $0.458 \pm 0.012$ \\
DASH & 20 & $0.774 \pm 0.005$ & $0.456 \pm 0.017$ \\
DASH & 0 & $0.773 \pm 0.002$ & $0.454 \pm 0.006$ \\
IRT & 0 & $0.771 \pm 0.007$ & $0.456 \pm 0.015$ \\
MIRTb & 20 & $0.770 \pm 0.007$ & $0.460 \pm 0.007$ \\
MIRTb & 5 & $0.770 \pm 0.004$ & $0.459 \pm 0.011$\\
PFA & 0 & $0.744 \pm 0.004$ & $0.481 \pm 0.004$ \\
AFM & 0 & $0.707 \pm 0.005$ & $0.499 \pm 0.006$ \\
PFA & 20 & $0.670 \pm 0.010$ & $1.008 \pm 0.047$ \\
PFA & 5 & $0.664 \pm 0.010$ & $1.107 \pm 0.079$ \\
AFM & 20 & $0.644 \pm 0.005$ & $0.817 \pm 0.076$ \\
AFM & 5 & $0.640 \pm 0.007$ & $0.941 \pm 0.056$ \\
\bottomrule
\end{tabular}
\caption{Performance comparison on the Algebra 2005-2006 (PSLC DataShop) dataset. Metrics are averaged over 5 folds and standard deviations are reported. $\uparrow$ and $\downarrow$ respectively indicate that higher (lower) is better.}
\label{algebra05_res}
}\qquad
\parbox{.42\linewidth}{%
\centering
\begin{tabular}{cccc}
\toprule
model &  dim & AUC $\uparrow$ & NLL $\downarrow$\\
\midrule
DAS3H & 5 & $\textbf{0.744} \pm 0.002$ & $\textbf{0.531} \pm 0.001$ \\
DAS3H & 20 & $\textbf{0.740} \pm 0.001$ & $\textbf{0.533} \pm 0.003$ \\
DAS3H & 0 & $\textbf{0.739} \pm 0.001$ & $\textbf{0.534} \pm 0.002$ \\
DASH & 0 & $0.703 \pm 0.002$ & $0.557 \pm 0.004$ \\
DASH & 5 & $0.703 \pm 0.001$ & $0.557 \pm 0.001$ \\
DASH & 20 & $0.703 \pm 0.002$ & $0.557 \pm 0.002$ \\
IRT & 0 & $0.702 \pm 0.001$ & $0.558 \pm 0.001$ \\
MIRTb & 20 & $0.701 \pm 0.001$ & $0.558 \pm 0.001$ \\
MIRTb & 5 & $0.701 \pm 0.002$ & $0.558 \pm 0.001$ \\
PFA & 5 & $0.669 \pm 0.002$ & $0.577 \pm 0.002$ \\
PFA & 20 & $0.668 \pm 0.002$ & $0.578 \pm 0.003$\\
PFA & 0 & $0.668 \pm 0.002$ & $0.579 \pm 0.002$ \\
AFM & 5 & $0.610 \pm 0.001$ & $0.597 \pm 0.001$ \\
AFM & 20 & $0.609 \pm 0.001$ & $0.597 \pm 0.003$ \\
AFM & 0 & $0.608 \pm 0.002$ & $0.598 \pm 0.002$ \\
\bottomrule
\end{tabular}
\caption{Performance comparison on the ASSISTments 2012-2013 dataset. Metrics are averaged over 5 folds and standard deviations are reported. $\uparrow$ and $\downarrow$ respectively indicate that higher (lower) is better.}
\label{assist12_res}
}
\end{table*}

We compared DAS3H to DASH, IRT, PFA, and AFM within the KTM framework, for three different embedding dimensions: 0, 5, and 20. When $d>0$, IRT becomes MIRTb, a variant of MIRT that considers a user bias. We do not compare to DKT, due to the mild controversy over its performance \cite{wilson2016back,wilson2016estimating}. For DASH, we choose to consider item-specific biases, and not KC-specific biases: in their original setting, Lindsey et al. \cite{lindsey2014improving} aggregated items into equivalence classes and trained DASH on this basis. This is not always possible to us because items have in general multiple skill taggings; however, we tested this possibility in Subsection \ref{discussion} but it did not yield better results.

\begin{sloppypar}
We used three different datasets: ASSISTments 2012-2013 (assist12) \cite{feng2009addressing},
Bridge to Algebra 2006-2007 (bridge06) and Algebra I 2005-2006 (algebra05) \cite{dataAlgebra}. The two latter datasets stem from the KDD Cup 2010 EDM Challenge. The main problem for our experiments was that only few datasets that combine both time variables and multiple-KC tagging are publicly available. As a result, only both KDD Cup 2010 datasets have items that involve multiple KCs at the same time. As a further work, we plan to test DAS3H on datasets spanning more diverse knowledge domains and having more fine-grained skill taggings. In ASSISTments 2012-2013, the \texttt{problem\_id} variable was used for the items and for the KDD Cup datasets, the item variable came from the concatenation of the problem and the step IDs, as recommended by the challenge organizers.
\end{sloppypar}
We removed users for whom the number of interactions was less than 10. We also removed interactions with NaN skills, because we feared it would introduce too much noise. For the KDD Cup 2010 datasets, we removed interactions which seemed to be duplicates, i.e. for which the (user, item, timestamp) tuple was duplicated. Finally, we sparsely encoded the features and computed the q-matrices. We detail the dataset characteristics (after preprocessing) in Table \ref{data_caracs}. The mean skill delay refers to the mean time interval (in days) between two interactions with the same skill, and the mean study period refers to the mean time difference between the last and the first interaction for each student.

\subsection{Results}

Detailed results can be found in Tables \ref{algebra05_res}, \ref{assist12_res} and \ref{algebra_res}, where mean area under the curve scores (AUC) and mean negative log-likelihood (NLL) are reported for each model and dataset. Accuracy (ACC) is not reported by lack of space. We found that ACC was highly correlated with AUC and NLL; the interested reader can find it on the Github repository containing code for the experiments\footnote{\url{https://github.com/BenoitChoffin/das3h}}. Standard deviations over the 5 folds are also reported. We can see that our model DAS3H outperforms all other models on every dataset.

\begin{figure}
   \caption{\label{AUC_benefit_all} AUC boost when using time windows features instead of regular wins and attempts (all datasets). Higher is better.}
    \includegraphics[width=\linewidth]{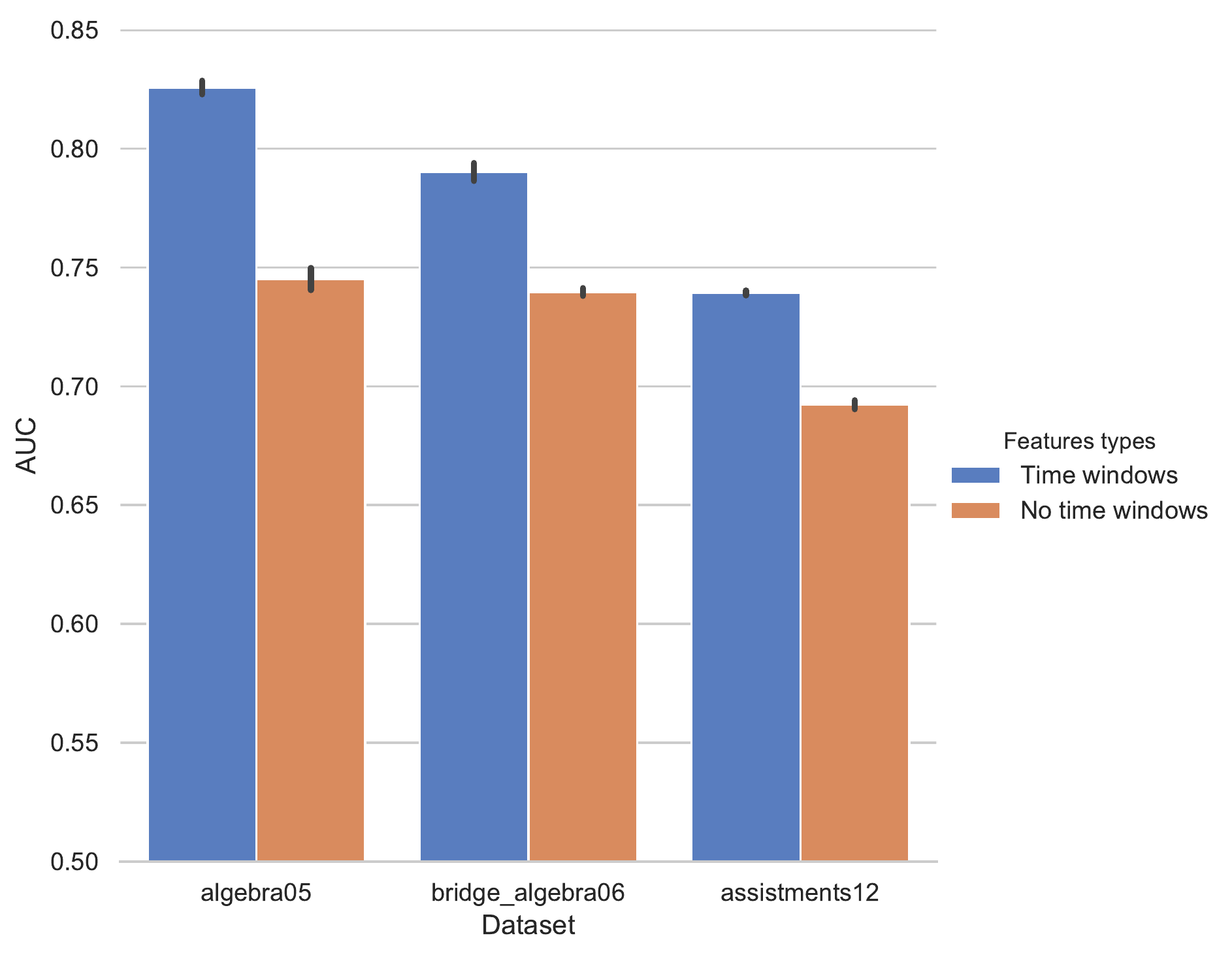}
   \vspace{-1cm}
\end{figure}

\subsection{Discussion}
\label{discussion}

Our experimental results show that DAS3H is able to more accurately model student performance when multiple skill and temporal information is at hand. We hypothesize that this performance gain stems from a more complex temporal modeling of the influence of past practice of skills on current performance.

The impact of the multidimensional embeddings and the pairwise interactions seems to be very small yet unclear, and should be further investigated. An embedding dimension of $d=20$ is systematically worse or among the worst for DAS3H on every dataset, but with a smaller $d=5$, the performance is sometimes better than with $d=0$. An intermediate embedding dimension could be preferable, but our results confirm those of Vie and Kashima \cite{Vie2019}: the role of the dimension $d$ seems to be limited.

In order to make more sense of our results, we wanted to know what made DAS3H more predictive than its counterparts. Our hypothesis was that taking the past temporal distribution of practice as well as the outcome of previous encounters with skills allowed the model to capture more complex phenomena than just simple practice, such as forgetting. To test this hypothesis, we performed some ablation tests. We empirically evaluated the difference in terms of AUC on our datasets when time windows features were used instead of regular features for wins and attempts. For each dataset, we compared the mean AUC score of the original DAS3H model with a similar model for which the time windows wins and attempts features were replaced with regular wins and fails counts. Thus, the time module $h_{\theta}$ was replaced with $\sum_{k \in KC(j)} \gamma_k c_{s,k} + \rho_k f_{s,k}$ like in PFA. Since wins, fails and attempts are collinear, it does not matter to replace ``wins and attempts'' with ``wins and fails''. The results are plotted in Figure \ref{AUC_benefit_all}. Mean and standard deviations over 5 folds are reported. We chose an embedding dimension $d=0$ since it was in general the best on the previous experiments. We observe that using time window features consistently boosts the AUC of the model.

We also wanted to know if assuming that skill practice benefits should differ from one skill to another was a useful assumption. Thus, we compared our original DAS3H formulation to a different version, closer to the DASH formulation, in which all skills share the same parameters $\theta_{2w+1}$ and $\theta_{2w+2}$ inside a given time window $w$. We refer to this version of DAS3H as $\textrm{DAS3H}_\textrm{1p}$. The results are given in Table \ref{comp_das3h_multiparam}. They show that using different parameters for different skills in $h_{\theta}$ increases AUC performance. The AUC gain varies between $+0.03$ and $+0.04$. This suggests that some skills have significantly different learning and forgetting curves.

One could argue also that this comparison between DAS3H and DASH is not totally accurate. In their papers, Lindsey et al. cluster similar items together to form disjoint knowledge components. This is not possible to perform directly for both KDD Cup datasets since some items have been tagged with multiple skills. Nevertheless, the ASSISTments 2012-2013 dataset has only single-KC items. To evaluate whether considering the temporal distribution and the outcomes of past practice on the KCs (DASH [KC]) or on the items (DASH [items]) would be better, we compared these two DASH formulations on ASSISTments 2012-2013. Detailed results can be found in Table \ref{comp_dash_items_kc}. We see that DASH [items] and DASH [KC] have comparable performance.

\begin{table}
\centering
\begin{tabular}{cccc}
\toprule
                      model &  dim &             AUC $\uparrow$ & NLL $\downarrow$\\
\midrule
DAS3H & 5 & $\textbf{0.791} \pm 0.005$ & $\textbf{0.369} \pm 0.005$ \\
DAS3H & 0 & $\textbf{0.790} \pm 0.004$ & $\textbf{0.371} \pm 0.004$ \\
DAS3H & 20 & $0.776 \pm 0.023$ & $0.387 \pm 0.027$ \\
DASH & 0 & $0.749 \pm 0.002$ & $0.393 \pm 0.007$ \\
DASH & 20 & $0.747 \pm 0.003$ & $0.399 \pm 0.002$ \\
IRT & 0 & $0.747 \pm 0.002$ & $0.393 \pm 0.007$ \\
DASH & 5 & $0.747 \pm 0.003$ & $0.399 \pm 0.002$ \\
MIRTb & 5 & $0.746 \pm 0.002$ & $0.398 \pm 0.006$ \\
MIRTb & 20 & $0.746 \pm 0.004$ & $0.399 \pm 0.007$ \\
PFA & 20 & $0.746 \pm 0.003$ & $0.397 \pm 0.004$ \\
PFA & 5 & $0.744 \pm 0.007$ & $0.402 \pm 0.007$ \\
PFA & 0 & $0.739 \pm 0.003$ & $0.406 \pm 0.008$ \\
AFM & 5 & $0.706 \pm 0.002$ & $0.411 \pm 0.004$ \\
AFM & 20 & $0.706 \pm 0.002$ & $0.412 \pm 0.004$ \\
AFM & 0 & $0.692 \pm 0.002$ & $0.423 \pm 0.006$ \\
\bottomrule
\end{tabular}
\caption{Performance comparison on the Bridge to Algebra 2006-2007 (PSLC DataShop) dataset. Metrics are averaged over 5 folds and standard deviations are reported. $\uparrow$ and $\downarrow$ respectively indicate that higher (lower) is better.}
\label{algebra_res}
\end{table}
\begin{table}
\centering
\begin{tabular}{ccccc}
\toprule
        &   $d$            &  bridge06 &             algebra05 & assist12 \\
\midrule
\parbox[t]{2.2mm}{\multirow{3}{*}{\rotatebox[origin=c]{90}{DAS3H}}} & 0 & $\textbf{0.790} \pm 0.004$ & $\textbf{0.826} \pm 0.003$ & $0.739 \pm 0.001$\\
& 5 & $\textbf{0.791} \pm 0.005$ & $0.818 \pm 0.004$ & $\textbf{0.744} \pm 0.002$ \\
& 20 & $0.776 \pm 0.023$ & $0.817 \pm 0.005$ & $0.740 \pm 0.001$ \\[1em]
\parbox[t]{2.2mm}{\multirow{3}{*}{\raisebox{-2ex}[0.2ex]{\rotatebox[origin=c]{90}{$\textrm{DAS3H}_\textrm{1p}$}}}} & 0 & $0.757 \pm 0.003$ & $0.789 \pm 0.009$ & $0.701 \pm 0.002$ \\
& 5 & $0.757 \pm 0.005$ & $0.787 \pm 0.005$ & $0.700 \pm 0.001$ \\
& 20 & $0.757 \pm 0.003$ & $0.789 \pm 0.006$ & 0.701 (<1e-3) \\[0.5em]
\bottomrule
\end{tabular}
\caption{AUC comparison on all datasets between DAS3H and $\textrm{DAS3H}_\textrm{1p}$, a version of DAS3H for which the influence of past practice does not differ from one skill to another. Standard deviations are reported. Higher is better.}
\label{comp_das3h_multiparam}
\end{table}
\begin{table}
\centering
\begin{tabular}{lccc}
\toprule
        DASH               &  $d=0$ &             $d=5$ & $d=20$ \\
\midrule                      
items & $0.703 \pm 0.002$ & $0.703 \pm 0.001$ & $0.703 \pm 0.002$ \\
KC & $0.702 \pm 0.001$ & $0.701 \pm 0.001$ & $0.701 \pm 0.001$ \\
\midrule
\end{tabular}
\caption{AUC comparison on ASSISTments 2012-2013 between DASH [items] and DASH [KC]. Standard deviations are reported. Higher is better.}
\label{comp_dash_items_kc}
\end{table}

Finally, let us illustrate the results of DAS3H by taking two examples of KCs of Algebra I 2005-2006, one for which the estimated forgetting curve slope is steep, the other one for which it is more flat. As a proxy for the forgetting curve slope, we computed the difference of correctness probabilities when a ``win'' (i.e. a correct outcome when answering an item involving a skill) left a single time window. This difference was computed for every skill, for every couple of time windows, and for every fold. The differences were then averaged over the 5 folds and over the different time windows, yielding for every skill the probability of correctness average decrease when a win leaves a single time window. One of the skills for which memory decays slowly concerns shading an area for which a given value is inferior to a threshold: in average and everything else being equal, the probability of correctness for an item involving this skill decreases by 1.15\% when a single ``win'' leaves a time window. Such a skill is indeed not difficult for a student to master with a few periodic reviews. On the contrary, the skill concerning the application of exponents is more difficult to remember as time goes by: for this KC, the correctness probability decreases by 2.74\% when a win leaves a time window. This is more than the double of the previous amount and is consistent with the description of the KC.

In brief, we saw in this section that DAS3H outperforms the other EDM models to which we compared it -- including DASH. Using time window features instead of regular skill wins and attempts counts and estimating different parameters for different skills significantly boosts performance. Considering that DAS3H outperforms its ablated counterparts and DASH, these results suggest that including both item-skill relationships and forgetting effect improves over models that consider one or the other. Using multidimensional embeddings, however, did not seem to provide richer feature representations, contrary to our expectations.

Besides its performance, DAS3H has the advantage to be suited to the adaptive skill practice scheduling problem we described in Section \ref{pb_form}. Indeed, it encapsulates an estimation of the current mastery of any skill and combination of skills for student $s$. It can also be used to infer its future evolution and thus, be leveraged to adaptively optimize a personalized skill practice schedule.

\section{Conclusion and future work}

In this article, we first formulated a research framework for addressing the problem of optimizing human long-term memory of skills. More precisely, the knowledge to be remembered here is \textit{applicative}: we intend to maximize the period during which a human learner will be able to leverage their retention of a skill they practiced to answer an item involving this skill. This framework assumes multiple skills tagging and is adapted to the more common flashcards-based adaptive review schedulers.

We take a student modeling approach to start addressing this issue. As a first step towards an efficient skill practice scheduler for optimizing human long-term memory, we thus propose a new student learning and forgetting model coined DAS3H which extends the DASH model proposed by Lindsey et al. \cite{lindsey2014improving}. Contrary to DASH, DAS3H allows each item to depend on an arbitrary number of knowledge components. Moreover, a bias for each skill temporal feature is estimated, whereas DASH assumed that item practice memory decayed at the same rate for every item. Finally, DAS3H is based on the recent Knowledge Tracing Machines model \cite{Vie2019} because feature embeddings and pairwise interactions between variables could provide richer models. To the best of our knowledge, KTMs have never been used with memory features so far. Finally, we showed that DAS3H outperforms several state-of-the-art EDM models on three real-world educational datasets that include information on timestamps and KCs. We showed that adding time windows features and assuming different learning and forgetting curves for different skills significantly boosts AUC performance.

This work could be extended in different ways. First, the additive form of our model makes it compensatory. In other terms, if an item $j$ involves two skills $k_1$ and $k_2$, a student could compensate a small practice in $k_1$ by increasing their practice in $k_2$. This is the so-called ``explaining away'' issue \cite{wellman1993explaining}. Using other non-affine models \cite{lan2016dealbreaker} could be relevant.
\newline
Following Lindsey et al. \cite{lindsey2014improving}, we used 5 time windows for DAS3H during our experiments: $\{1/24,1,7,30,+\infty\}$. Future work could investigate the impact of alternative sets of time windows -- for instance, with more fine-grained time scales. However, one should pay attention not to add too many parameters to estimate.
\newline
Future work should also compare DAS3H and DASH to additional student models. For instance, R-PFA \cite{galyardt2015move} (Recent-Performance Factor Analysis) and PFA-decay \cite{gong2011construct} extend and outperform PFA by leveraging a representation of past practice that puts more weight on more recent interactions. However, they do not explicitly take the temporal distribution of past practice to predict future student performance. Other memory models, such as ACT-R \cite{pavlik2008using} or MCM \cite{pashler2009predicting} could also be tested against DAS3H.
Latency, or speed of recall, can serve as a proxy of retrieval difficulty and memory strength \cite{mettler2016comparison}. It would be interesting to test whether incorporating this information inside DAS3H would result in better model performance.
\newline
In a real-world setting, items generally involve multiple skills at the same time. In such a situation, how should one select the next item to recommend a user so as to maximize their long-term memory? The main issue here is that we want to anchor skills in their memory, not specific items. We could think of a two-step recommendation strategy: first, selecting the skill $k^\star$ whose recall probability is closest to a given threshold (this strategy is consistent with the cognitive psychology literature, as Lindsey et al. recall \cite{lindsey2014improving}) and second, selecting an item among the pool of items that involve this skill. However, it could be impossible to find an item that involves \textit{only} this skill $k^\star$. Also, precocious skill reactivations can have a harmful impact on long-term memory \cite{cepeda2008spacing}. Thus, a strategy could be to compute a score (weighted according to the recall probability of each individual skill) for each skill combination in the q-matrix and to choose the combination for which the score is optimized.
\newline
Finally, we tested our model on three real-world educational datasets collected from automatic teaching systems on mathematical knowledge. To experiment with our model, we were indeed constrained in our choice of datasets, since few publicly available of them provide both information on the timestamps and the skills of the interactions. As further work, we intend to test our model on other datasets, from more diverse origins and concerning different knowledge domains. Collecting large, fine-grained and detailed educational datasets concerning diverse disciplines and making them publicly available would more generally allow EDM researchers to test richer models.

\section*{Acknowledgements}

We would like to warmly thank Pr. Mozer from Univ. of Colorado, Boulder for providing useful details on their papers \cite{mozer2016predicting, lindsey2014improving} and allowing us to access the data of their experiment, and to Alice Latimier (LSCP, Paris) for her crucial comments concerning the cognitive science part. This work was funded by Caisse des Depôts et Consignations, e-Fran program.

\fontsize{9.0pt}{10.0pt}\selectfont
\bibliographystyle{abbrv}
\bibliography{mybiblio}
\balancecolumns
\end{document}